# Influence of short time milling in $R_5(Si,Ge)_4$, R =Gd and Tb, magnetocaloric materials

A.L. Pires[1,2], J.H. Belo[1], J. Turcaud[3], G.N.P. Oliveira[1,2], J.P. Araújo[1],
A. Berenov[3], L. F. Cohen[3], A.M.L. Lopes[1,2], A.M. Pereira[1,3*]

[1]*IFIMUP and IN - Institute of Nanoscience and Nanotechnology, Departamento de Física e Astronomia da Faculdade de Ciências da Universidade do Porto, Rua Campo Alegre, 687, 4169-007 Porto, Portugal*

[2]*CFNUL - Centro de Física Nuclear da Universidade de Lisboa, Av. Prof. Gama Pinto, 2, 1649-003 Lisboa, Portugal*

[3]*Blackett Laboratory, Imperial College, London SW7 2AZ*

The effect of the short milling times on $R_5(Si,Ge)_4$ R =Gd, Tb magnetocaloric material properties was investigated. In particular, the effect of milling on atomic structure, particles size and morphology, magnetic, and magnetocaloric effect was studied. With short milling times (< 2.5h), a reduction of the $Gd_5Si_{1.3}Ge_{2.7}$ and $Tb_5Si_2Ge_2$ particles size was achieved down to approximately 3.5 μm. For both compositions the main differences are a consequence of the milling effect on the coupling of the structural and magnetic transitions. In the $Gd_5Si_{1.3}Ge_{2.7}$ case, a second-order phase transition emerges at high temperatures as a result of ball milling. Consequently, there is a decrease in the magnetocaloric effect of 35% after 150 minutes of milling. Interestingly, an opposite effect is observed in $Tb_5Si_2Ge_2$ where a 23% increase of the magnetocaloric effect was achieved, driven by the enhancement of the coupling between magnetic and structural transitions arising from internal strain promoted by the milling process.

## I. INTRODUCTION:

Magnetic cooling systems at room temperature have emerged as an alternative to the conventional refrigeration technology due to two breakthrough discoveries at the end of the 1990's: the magnetic refrigerator prototype developed by Zimm *et al* [1] in 1998 and the Giant Magnetocaloric Effect (GMCE) discovery by V. Pecharscky and K. Gschneider reported in 1997 [2]. These milestones led to a remarkable increase in the number of room temperature magnetic refrigerator prototypes developed per year [3,4]. Concerning prototyping, Zimm and co-workers showed a cooling power of 600W and temperature variation of 10K, depending on the amount of the magnetocaloric material used (gadolinium spheres) [1,4,5]. This achievement promoted an exponential interest in the search for optimized magnetocaloric materials because it offered the possibility of using green energy, compared to today's conventional cooling systems such as air conditioners and freezers [6]. One of the key points of this technology is the magnetic material, since the GMCE is associated with a first-order transition, arising from the simultaneous occurrence of both magnetic and structural transitions [7]. Till now, several families of compounds exhibiting the magnetocaloric effect (MCE) near room temperature were discovered, such as: $Fe_{0.49}Rh_{0.51}$, $Gd_5Si_2Ge_2$ [2,8,9], $La(Fe,Si)_{13}$ [10–12]; Ni-Mn-In(Co) [13–15] and MnAs [16]. Since their discovery, there has been a constant effort to understand how to improve the magnetocaloric properties of each family [17]. Also several of these materials belong to the group of the magnetic shape memory alloys that display magnetic field-induced structural transitions [18,19]. Materials that belong to the $R_5(Si,Ge)_4$ family, are among the most attractive candidates for magnetic refrigeration systems, since these materials are quite resistant to corrosion in the air/water and present a GMCE over a wide range of temperatures [20]. As the prototype development and optimization continues, other challenges have appeared such as the need to reduce the heat exchange time, as well as diminish the undesirable magnetic hysteresis loss [8,21–24]. One strategy is to tune the chemical stoichiometry in order to tailor the magnetic transition to the region at the border between the first and second-order type of transition (known as the critical point) [25]. Moreover, equally important is the design of the material geometry and in particular the size reduction of these materials in order to increase their surface-area/volume ratio [26]. The size reduction also allows for flexibility in design and enhanced functionalities. The reduced size [27,28] usually enhances the contribution of surface effects, leading to changes to the MCE itself [29,30]. On the $R_5(Si,Ge)_4$ compounds, Trevizoli *et al* [26], showed that by using pulverization powder metallurgy technique it is possible to tailor the particles size distribution and influence the magnetic properties of the $Gd_{5.09}Ge_{2.03}Si_{1.88}$ alloy. In that work, residual tension that is formed during the sintering has a direct influence on the suppression of the first-order phase transition. In fact, the first-order phase transition is eliminated during sintering in samples with the lowest particle sizes (between 32 and 53 µm). Instead, bigger particles (106 and 149 µm) retain a coupled magnetostructural transition with a large magnetic entropy value of -14.0 $JKg^{-1}K^{-1}$ [26].

* Corresponding author: Dr. André Pereira, Tel.: +351 220402369; fax: +351220402406.
*E-mail address:* ampereira@fc.up.pt;

Ball milling (BM) is one of the most attractive routes to scale-down material size to the micrometer length scale [31,32]. Rajkumar *et al* [33] originally presented a work using ball milled powders, reaching the 1-5 µm size on $Gd_5Si_2Ge_2$ and $Gd_5Si_2Ge_{1.9}Fe_{0.1}$ alloys. They observed that for the sample with Fe content, 32 h of BM promotes a splitting of the bulk magnetic transition towards two different magnetic transitions. Furthermore, a drastic decrease in the magnetic entropy change value (to 0.45 $JKg^{-1}K^{-1}$, with ΔH=2T) in $Gd_5Si_2Ge_2$ powders was observed. In this case, the particle size reduction is associated with a decrease of the magnetic moment, suggesting the randomization of the magnetic moment at the surface [33]. Others studies reported the influence of the particle size decrease on the magnetocaloric effect [34–37]. Phan et al [34], reported the comparison of iron garnets nanoparticles ($Gd_3Fe_5O_{12}$) produce by BM and the bulk counterpart. In their study the authors report that an increase of the MCE is observed when the grain size is decreased to 35nm. The BM samples show X-ray diffraction peak broadening effects, associated with the induce strain and introduction of defects into the lattice. Also, Biswas et al [36] also found an enhancement of the MCE and suggested that this was related to the ordering of surface spins that is more evident at low temperatures. Thus, Eu8Ga16Ge30 shows two transitions (first and second-order transition) and both transitions show different behavior depending on the crystallite size produced by BM. The second-order transition shows an increase of MCE with the decrease of crystallite size, and the first-order transition the opposite effect. It is thought that the reduction of crystallite size promotes the increase of the fraction of non-crystalline grain boundaries that results in the disruption of the first order ferromagnetic ordering [36]. Recently, Giovanna do Couto et al [37] reported a higher value of magnetocaloric effect also using the BM process, where the main difference from the Rajkumar *et al.* work, was the shorter milling time (~ 4h). Nevertheless, Giovanna do Couto and co-authors reached lower mean particles size (0.5 µm for $Gd_5Si_2Ge_2$) and obtained a – $\Delta S_M$ ~ 4 $JKg^{-1}K^{-1}$ (with $\Delta H$=0-5 T), corresponding to an 80% reduction when compared with bulk counterpart value of 20 $JKg^{-1}K^{-1}$. Again, the reduction was related to the suppression of the amount of sample undergoing a first-order transition [37], namely by its transformation into an amorphous phase due to the increased milling time. Thus, in the context of the previous studies, the aim of the present work is to perform a systematic study of the effect of shorter milling times (<2.5h) on atomic structure, morphology, magnetic and MCE and to monitor the effect of that process in detail. The exploration of the MCE with the reduction of grain size relative to the bulk materials is the focus of this work. In order to understand the role played by the shorter BM times on the magnetic and/or structure transition and their (de)coupling, two compositions were studied: $Gd_5Si_{1.3}Ge_{2.7}$ and $Tb_5Si_2Ge_2$ alloys. The $Gd_5Si_{1.3}Ge_{2.7}$ compound presents a simultaneous magnetic and structural (magnetostructural) transition at 181 K whereas $Tb_5Si_2Ge_2$ exhibits a shift of a few Kelvins between the magnetic ($T_C$) and the structural transition ($T_S$) ($T_S$~105 K and $T_C$~112 K on heating) [38,39].

**II. EXPERIMENTAL DETAILS**

The $Gd_5Si_{1.3}Ge_{2.7}$ and $Tb_5Si_2Ge_2$ buttons were prepared in the arc melted furnace by mixing stoichiometric amounts of commercial pure rare earth elements (Gd or Tb - 99.999% Wt) with Si (99.9995% Wt) and Ge (99.999% Wt) in an argon atmosphere. After the initial melting, three re-melts were performed to improve the sample homogeneity. These buttons were ground using the BM technique, with the purpose of reducing the particle size. Therefore, the buttons (previously lightly fragmented) were placed in a steel vessel with zirconia balls (~ 12 spheres) in a planetary BM configuration. The sample, with 3-4 g of powder, was milled in isopropanol with fixed rotation (400 rpm). For $Gd_5Si_{1.3}Ge_{2.7}$ the mill time was varied from 30, 90, 120 and 150 minutes and for $Tb_5Si_2Ge_2$ 60, 90, 120 and 150 minutes. Finally, the vessel containing the material and ethanol was placed in a stove at 60 °C for 12 hours to evaporate the isopropanol.

The structural characterization of the milled powders was evaluated from the X-ray diffraction patterns (XRD) obtained using a *Panalytical, X´Pert Pro* diffractometer with a CuKα radiation and λ=1.54056Å at 40 kV and 30 mA. The crystallographic structures were determined by Rietveld refinements of the XRD spectrum, using the FULLPROF software package [40]. Such refinements were performed in a systematic way by considering the presence of two different crystallographic phases (in the case of the $Gd_5Si_{1.3}Ge_{2.7}$), which were added into the refinement one at a time in order to minimize errors in peaks assignment. The microstructure and chemical composition of the samples with different milling times were evaluated by electron microscopy (SEM) using a *Philips-FEI/Quanta 400* with a coupled energy dispersive spectroscope (EDS). The magnetization curves, i.e. the temperature dependence of the magnetization (M(T) curves) and isothermal magnetization curves (M(H)) were measured in a Quantum Design Physical Property Measurement System (PPMS) with a Vibrating Sample Magnetometer setup option in the case of the $Gd_5Si_{1.3}Ge_{2.7}$ and in a commercial (MPMS Quantum Design) Superconducting Quantum Interference Device (SQUID) magnetometer in the [100; 300] K temperature range, for $Tb_5Si_2Ge_2$. The M(H) magnetic isotherms curves were measure up to 50 kOe. Afterwards, the magnetic entropy changes [-ΔSm (T)] were estimated through the application of the Maxwell relation, according to the loop method, as report in Ref. [41].

## III. RESULTS AND DISCUSSION

### A. Structural Characterization

The effects of short milling times on the crystallography were evaluated in the two magnetocaloric materials: $Gd_5Si_{1.3}Ge_{2.7}$ and $Tb_5Si_2Ge_2$. In Fig. 1a) and 1b) the XRD patterns of $Gd_5Si_{1.3}Ge_{2.7}$ and $Tb_5Si_2Ge_2$ bulk and powders with different milling times are shown. The bulk patterns are characterized by sharp and well defined peaks, signaling high crystallinity. The milling process promotes the decrease/disappearance of some peaks. In case of $Gd_5Si_{1.3}Ge_{2.7}$, with increasing milling time, it is clearly observed that the following peaks intensities decrease: 30.56º [3 0 2]; 31.15º [3 1 -2]; 31.95º [3 2 1]; 32.70º [3 1 2]; 33.43º [4 2 0]; 33.92º [5 1 -1] and 35.01º [4 2 -1]. These peaks are characteristic of the monoclinic (M) phase [42]. After 150 min of BM some peaks appear that are characteristic of orthorhombic phase (O(I)),: 31.60º [1 3 2]; 32.93º [2 0 2] and 35.64º [1 4 2] [42,43]. Furthermore, all peaks show a slight displacement suggesting that volume changes are occurring, but this will be discussed further below. In order to extract the amount of M vs O(I), phases and structural properties, in $Gd_5Si_{1.3}Ge_{2.7}$ samples the patterns have been refined considering two crystalline phases, a M structure ($Gd_5Si_2Ge_2$ structure, $P112_1/a$ space group) and a O(I) structure ($Gd_5Si_4$ structure, Pnma space group). From the refinements, a linear reduction of volume of both phases with increase of the BM time up to 1% is obtained, as can be observed in Fig. 1. This feature explains the shifting of the XRD pattern peaks.

In accordance with this shrinking, a systematic increase of O(I) is obtained at room temperature increasing up to 40% for 150 minutes (see inset Fig. 2). In addition the XRD line broadening is show in Fig. 1a) and it is found to be proportional to the milling time. These results are in agreement with other studies like: $Gd_3Fe_5O_{12}$ and $Eu_8Ga_{16}Ge_{30}$ clathrate nanocrystals [34,36].

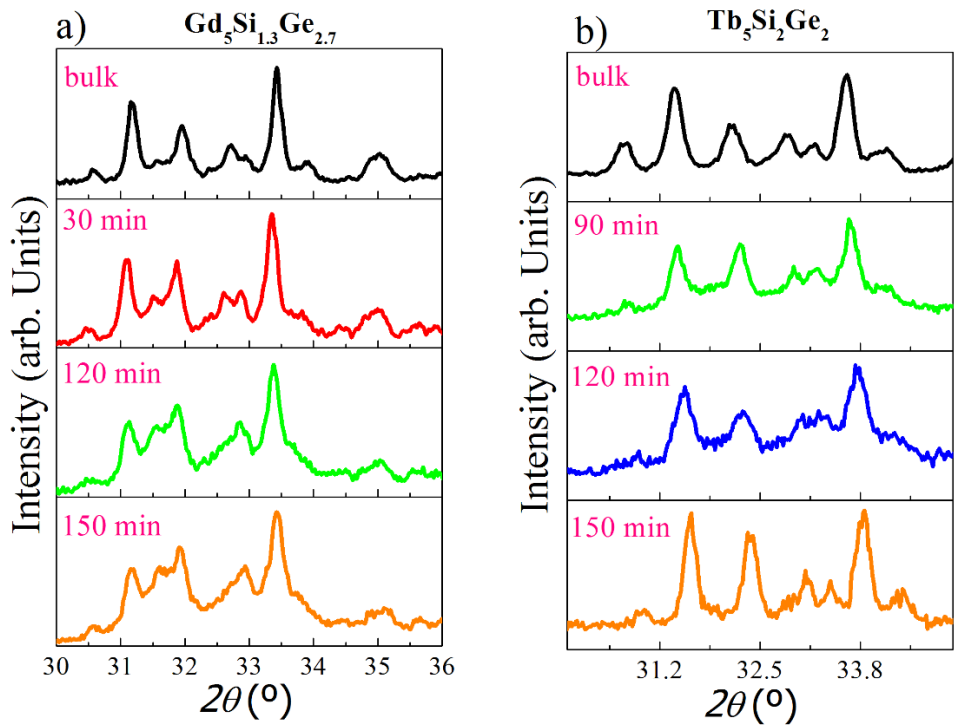

Figure 1 – (color online) XRD patterns of the powders with different millings times for: a) $Gd_5Si_{1.3}Ge_{2.7}$ (blue region corresponds to M reflections, the violet corresponds to reflections characteristic of the O(I) phase) and b) $Tb_5Si_2Ge_2$ (the dashed lines are a guide-to the-eye only).

In the case of $Tb_5Si_2Ge_2$, the XRD pattern remains almost unaltered demonstrating that at room temperature the stable 5:4 phase is the M phase. Nevertheless, an evident shift of the peaks to higher angles is observed. Like in the Gd case, these shifts are related to a volume unit cell decrease that were quantified in Rietveld refinements and depicted in Fig. 2, showing that BM time acts on both materials in a similar way, shrinking them. In the literature, these changes, are usually associated with structural disorder (vacancies, dislocations and grain boundaries) promoted by the milling process [31,37,44]. But in the present case it is clear that the peak shifts are related to strain effect induced by the BM [44].

One should mention that no $ZrO_2$ peaks, typically found in materials prepared by the milling process, are observed in our XRD patterns. This common contamination, absent in our work, results from the zirconium balls used in the milling process [37]. This shows one advantage of using short milling times and only 400 rpm in the milling process, contrary to what happens for longer milling times in other reports [37].

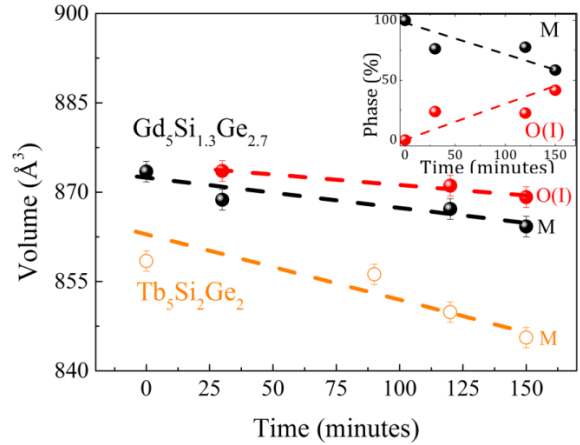

Fig. 2 – (color online) The volume as a function of time for $Gd_5Si_{1.3}Ge_{2.7}$ and $Tb_5Si_2Ge_2$ samples. Inset: the amount of the M and O(I) structure presented in $Gd_5Si_{1.3}Ge_{2.7}$ obtained by Rietveld refinement. Dashed lines are guides for the eyes.

### B. Particle Size and Morphology

Fig. 3 shows the images obtained by scanning electronic microscopy (SEM) that were used to analyze size and morphology of the $Gd_5Si_{1.3}Ge_{2.7}$ powder resulting for the milling process. Fig. 3a) to c) depicts the change in micromorphology of the powder resulting from different milling times: 30, 90 and 150 min.

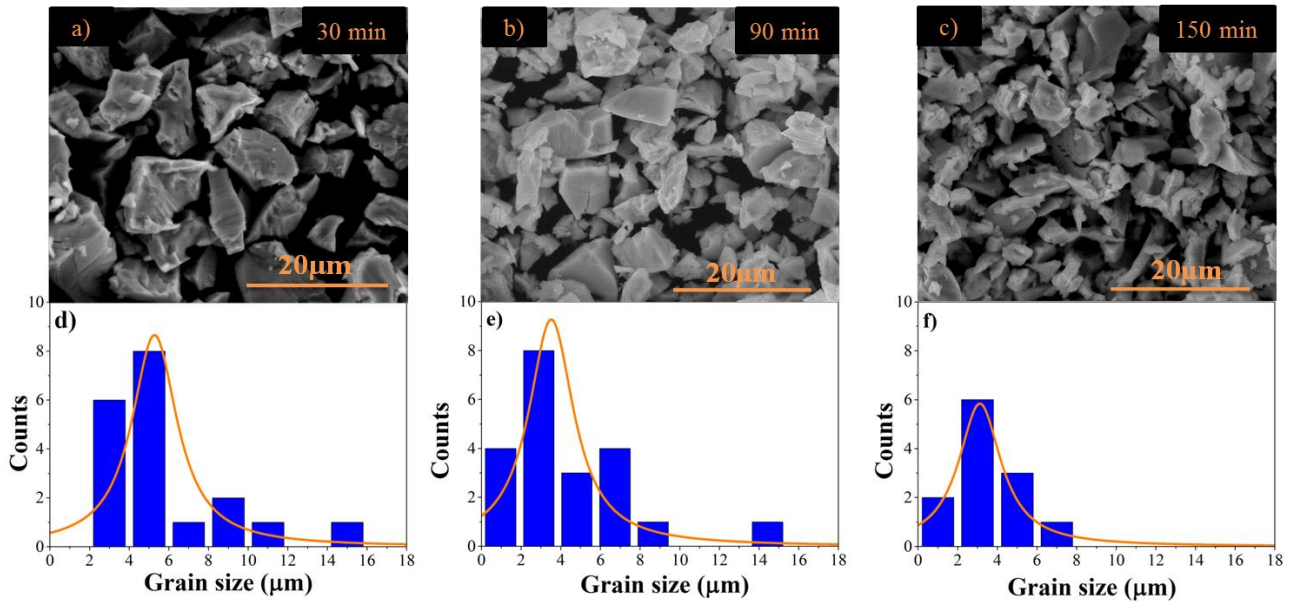

Fig. 3 – (color online) Scanning electron micrographs of $Gd_5Si_{1.3}Ge_{2.7}$ with different milling times: a) 30, b) 90 and c) 150 minutes; and the corresponding histograms: d), e) and f).

The micrographs show that the particles lack a uniform size and do not have a well-defined geometric morphology. By averaging across the images the mean particle size was estimated as 5.7, 4.5 and 3.6 µm for 30, 90 and 150 min milling times, respectively. These values were extracted from the fitted size distribution, shown in Fig. 3d) to f), using a Lorentzian function. Basically, the micrographs reveal that: 1) a decrease in the average particle size with the increase of the milling time due to fracturing; and 2) for milling times > 90 minutes, the particles form agglomerates that are associated with the higher reactivity of the powder as a result of the increase of the area/volume ratio of the particles. The agglomerate formation is very common with this technique [33,44]. In comparison with previous work by Giovanna do Couto et al [37], these results show a smaller distribution in particle size for the same 120 min of milling process with 400 rpm. This demonstrates the importance of the initial button grinding, before introducing it in the planetary. Similar results are achieved in SEM images of the in $Tb_5Si_2Ge_2$ and to simplify these will be not shown.

### C. Magnetic Characterization

The magnetization was measured as a function of temperature with constant magnetic field (H=1000 Oe) for both the $Gd_5Si_{1.3}Ge_{2.7}$ and $Tb_5Si_2Ge_2$ alloys, with the objective of studying the effect of the milling times on the magnetic properties. The temperature dependence of the magnetization for $Gd_5Si_{1.3}Ge_{2.7}$ samples are shown in Fig. 4a). In the case of $Tb_5Si_2Ge_2$ sample, since the structural and magnetic transition occurs at different but closely spaced temperatures, the (de)coupling is more easily observed in the temperature derivative (dM/dT) displayed in Fig. 4b.

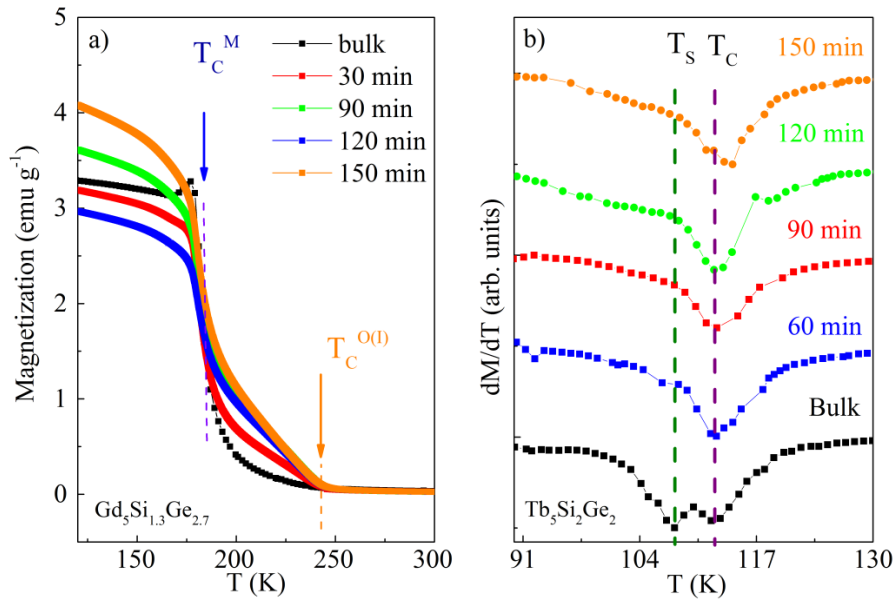

Fig. 4 – (color online) Temperature dependence of the magnetization for a) $Gd_5Si_{1.3}Ge_{2.7}$ and b) derivative of the M(T) curve for $Tb_5Si_2Ge_2$ with different milling times. Dashed lines are guides for the eyes.

For the $Gd_5Si_{1.3}Ge_{2.7}$ bulk sample, in the 120-300 K temperature range (on heating), Fig. 4a), a single sharp transition is observed at 181 K which is associated, as in previous works, with first-order transition where simultaneous magnetic and structural transition occurs, from orthorhombic I, [O(I)], – ferromagnetic to monoclinic, [M], – paramagnetic [8,45]. For the milled samples, it is clear that this process promotes significant changes in the magnetic properties when compared with the bulk sample, and corroborates the previous structural characterization: the decrease of the M phase responsible for the smoothening of the ~181 K transition, and the increase the O(I) phase responsible for the enhancement of the ~ 236 K transition. In fact, besides presenting the low temperature transition, which is present in the bulk sample, an additional second magnetic transition is apparent at ~236 K. From the literature, it appears that this transition is associated with a pure magnetic transition from paramagnetic to ferromagnetic promoted by the arrested of the O(I) structure up to room temperature and thus corresponds to a transition with second-order character. With increased BM time, the second-order transition becomes more pronounced revealing an increase of the O(I) structure at room temperature.

On the other hand the $Tb_5Si_2Ge_2$ bulk sample presents two transitions visible from the dM/dT curve, both within the temperature range of 91-130 K, on heating (Fig. 4b). The one at 108 K is associated with the structural transition ($T_s$) and the one at 112 K is associated with the magnetic transition ($T_C$), as reported in literature [46]. The structural transition occurs from O(I) → M phase, which is in accordance with our previous reports [39]. At higher temperatures (112 K), the transition is related to the change from ferromagnetic to paramagnetic state, while preserving the M structure. With increased milling time it is observed that the lower temperature transition starts to be less clear and even disappears completely for the longer milled sample (at 150 min). Simultaneously, there is a slightly shift in the high temperature transition, increasing by a few degrees (from 112 K to 114 K).

### D. Magnetic Entropy

The magnetic entropy variation with the milling time was performed in all the samples, and is presented in Fig. 5a) for the $Gd_5Si_{1.3}Ge_{2.7}$ and b) for $Tb_5Si_2Ge_2$ for magnetic field of 50 kOe.

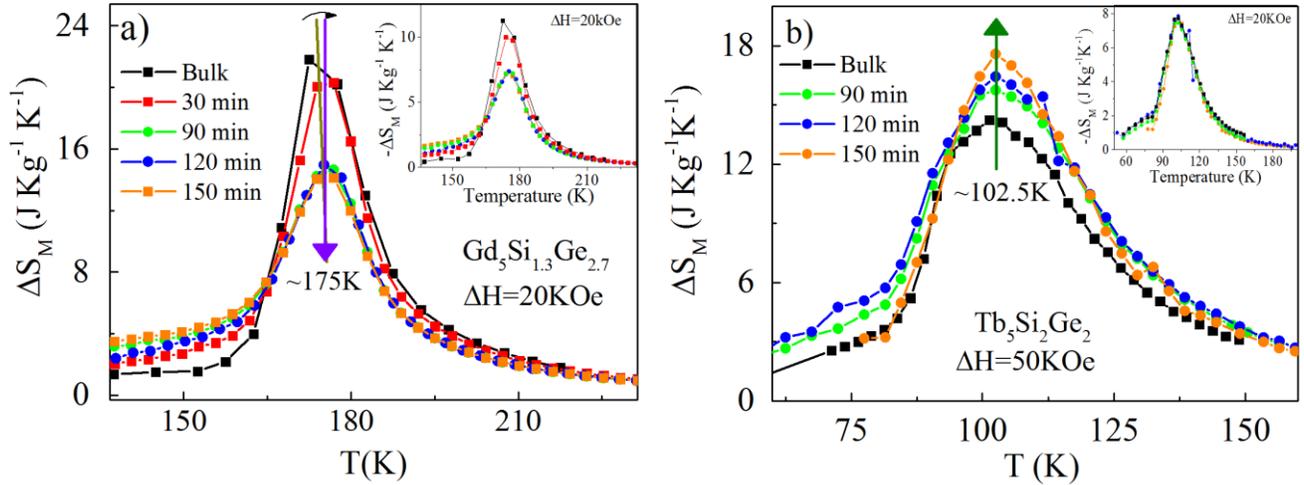

Fig. 5 – (color online) Temperature dependence of the magnetic entropy change for a) $Gd_5Si_{1.3}Ge_{2.7}$ and b) $Tb_5Si_2Ge_2$ and the effect of different millings times, under an applied magnetic field change of ΔH=50KOe (insets for reduced magnetic fields ΔH=20KOe).

From Fig. 5 it is observed opposite effects of milling times on $-\Delta S_M$ for the different compositions, i.e. milling decreases $-\Delta S_M$ for $Gd_5Si_{1.3}Ge_{2.7}$ and increases $-\Delta S_M$ for $Tb_5Si_2Ge_2$. This is more clearly observed when plotting the maximum value of the magnetic entropy change ($-\Delta S_M^{max}$) as a function of milling times for a) $Gd_5Si_{1.3}Ge_{2.7}$ and b) $Tb_5Si_2Ge_2$, as shown in Fig. 6. In the case of the $Gd_5Si_{1.3}Ge_{2.7}$, the bulk sample exhibits a $-\Delta S_M^{max}$ value of 21.8 $JKg^{-1}K^{-1}$ for ΔH=50 KOe. After 30, 90, 120 and 150 min of milling there is a $-\Delta S_M^{max}$ decrease of 8%, 33%, 32% and 35%, respectively. So, for $Gd_5Si_{1.3}Ge_{2.7}$ the magnetocaloric effect decreases when the milling time increases, however for longer times (>90 min) this parameter tends to a plateau (see Fig. 6a)). Note that there is a small shift of temperature at which the peak of the MCE occurs and that this peak tends to stabilize at 175 K. In summary, the milling process leads to a reduction of the maximum magnetic entropy change in agreement with previously reported data [44].

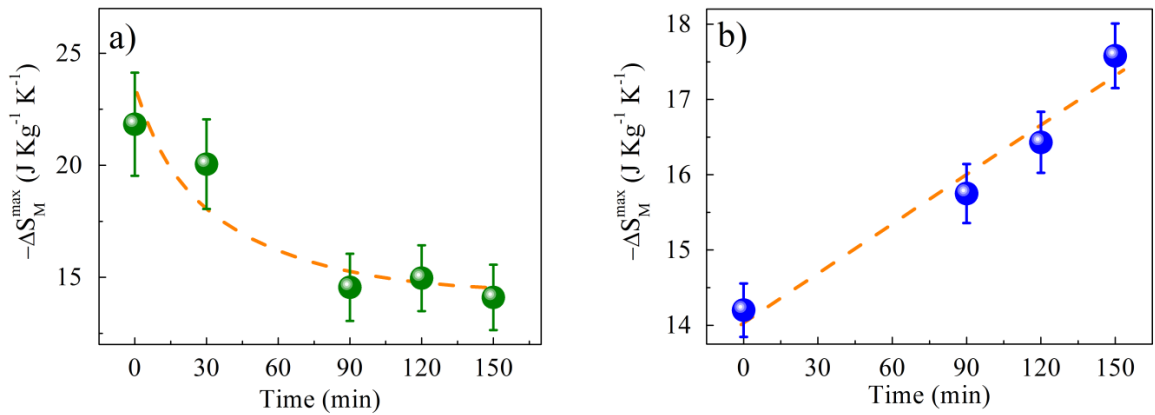

Fig. 6 – Maximum value of magnetic entropy change ($\Delta S_M^{max}$) as a function of milling times for a) $Gd_5Si_{1.3}Ge_{2.7}$ and b) $Tb_5Si_2Ge_2$. Dashed lines are guides for the eyes.

Fig. 6b shows the $-\Delta S_M(T)$ for the $Tb_5Si_2Ge_2$ bulk sample and this sample has a $-\Delta S_M^{max}$ value of 14.2 $JKg^{-1}K^{-1}$, for $\Delta H=50$ KOe, close to the value reported by Morellon and co-authors [47]. After 90, 120 and 150 min there is $-\Delta S_M^{max}$ increase of 11%, 16% and 23%, respectively. The temperature of the maximum magnetic entropy change is ~102.5K and remains constant for all milling times. The $Gd_5Si_{1.3}Ge_{2.7}$, $-\Delta S_M^{max}$ tends to stabilize with the increase of the milling time (in particular after 90 min), whereas the $Tb_5Si_2Ge_2$ shows a linear increase with the milling times studied in this work.

## IV. DISCUSSION

From the present work it is clear that using the same BM conditions produces opposite behavior in the otherwise similar compounds (in terms of their magnetic behaviour and atomic structure).

In the $Gd_5Si_{1.3}Ge_{2.7}$ case, the results follow the same trend as previous reported for $Gd_5Si_{1.8}Ge_{1.8}Sn_{0.4}$ [20] and $Gd_5Si_2Ge_{1.9}Fe_{0.1}$ [33]. In those works a suppression of the first-order transition (partial or total) was observed with BM. Most similar to our case is the report of magnetization behavior of $Gd_5Si_2Ge_{1.9}Fe_{0.1}$ where the change of magnetization was attributed to the suppression of the M structural phase and hence of its phase transition towards the O(I) structure [18,33,48,49]. Accordingly to the literature, it is suggested that the origin for the observed phenomena was related to: 1) particle size reduction; 2) strain/defect formed during the milling process; 3) chemical inhomogeneity or/and 4) compositional fluctuations. On the other hand, Giovanna do Couto et al [37] reported a detailed BM study where even for the shortest time considered (which was for 4 h) a quite drastically reduction of the first-order transition in $Gd_5Si_2Ge_2$ was observed. To explain the obtained properties the authors used similar explanation as the one proposed by Rajkumar and co-authors but adding that the different particle sizes also play a role on the magnetic properties because the downsizing implies changes in the magnetic interactions leading to the appearance of surface effects such spin canting/frustration arising from uncompensated spin interactions at the atomic surface [33]. Nevertheless, the difference that distinguishes this work is that by examining much shorter BM time we were able to monitor with detail the structural and magnetic properties changes. This provides insight concerning the role played by BM namely, that first hours of milling changes the internal structure of the materials by inducing internal strain that is able to change M phase into O(I) structure, reducing the amount of the first-order phase responsible for giant MCE whilst reducing the particle size. Furthermore, during the short time BM we were not able to detect the presence of defects and impurities from the XRD patterns and stoichiometry analysis. Only for longer times would we expect a drastic reduction of the crystallinity of the samples leading to the suppression of a magnetic transition and poorer or even null MCE values. As a result for short BM times we have achieved quite impressive MCE values. Table 1 summarizes the ΔSM values obtained for various ball milled $R_5(Si,Ge)_4$ materials reported in the literature namely for the Gd samples.

Table 1 – Review of the properties and the effect of BM in the $Gd_5(Si,Ge)_4$.

| Sample | Milling Time (min) | Grain Size (μm) | $T_C$ | $\Delta S_M^{max}$ (J/Kg K) | Ref. |
|---|---|---|---|---|---|
| $Gd_5Si_2Ge_2$ | 1920 | 1-5 | 225 | 0.45 | [33] |
| $Gd_5Si_2Ge_2$ | 0 | - | 269 | 20 | [37] |
|  | 240 | 0.5 | 261 | ~4 | [37] |
| $Gd_5Si_{1.3}Ge_{2.7}$ | 0 | - | 181 | 21.8 | In this work |
|  | 150 | 3.59 | 180 & 236 | 14.6 | In this work |

Here we show that BM in $Tb_5Si_2Ge_2$ compound induces an increase of 23% of the maximum magnetic entropy change. Still presenting an opposite effect, the physical explanation is similar to the Gd sample, i.e. induced strain with BM. It is well known that by applying hydrostatic pressure a full coupling of the magnetic and structural transition can be obtained in these compounds [47,50]. Thus, leading to a progressive increase of $\Delta S_M^{max}$ value from 13.4 JKg$^{-1}$K$^{-1}$ (P=0 kbar) till 22.1 JKg$^{-1}$K$^{-1}$ (P=9 kbar), for a 50 kOe magnetic field. This is analogous to that obtained in our results since the $\Delta S_M^{max}$ increase linearly up to ~17 JKg$^{-1}$K$^{-1}$ for 120 min milling. Furthermore in a previous work [50], our group reported similar effect when Fe is substituted in $Tb_5Si_2Ge_2$ which lead to an increase of the MCE, caused by internal strains in the grain boundaries that are enhanced by the Fe presence. Other materials report the same enhancement of the MCE with the decrease of grain size by BM. For example, nanostructures of iron garnets with 35 nm show an increase of the $\Delta S_M^{max}$ from 2.45 JKg$^{-1}$K$^{-1}$ to 3.47 JKg$^{-1}$K$^{-1}$ with $\Delta\mu_0 H$=3T [34]. $Eu_8Ga_{16}Ge_{30}$ clathrate (type-I), with average crystal size of 42 nm, shown an enhancement of the –ΔSM, to ~10 JKg$^{-1}$K$^{-1}$ for 5T, with decreasing crystal size (~15 nm), at lower temperatures.

Analyzing the ΔSM for low fields different behaviors depending on the materials were observed. $Gd_5Si_{1.3}Ge_{2.7}$ shows similar behavior for low and high applied magnetic fields, i.e. a –ΔSM decrease occurs with milling time. Such result indicates that even for ΔH=20kOe, the magnetic field is strong enough to promote the MS transition. In contrast, for $Tb_5Si_2Ge_2$ the ΔH= 20 kOe proved to be insufficient to promote a complete structural transition in the ball milled samples. Consequently, there are no significant changes in between the –ΔSM (T) curves for the ball milled samples – their differentiating factor is obvious in low applied fields. All the materials were measured in powder form with similar particle shape which allows us to neglect stray field effects since the demagnetization factor will be the same [51].

## IV. CONCLUSION

In this work we show that short mill times produce different results on magnetic properties depending on the type of the material processed. In particular, in $Tb_5Si_2Ge_2$ milling appears to promote stronger coupling of the structural and magnetic transition in contrast to $Gd_5Si_{1.3}Ge_{2.7}$. The SEM images show, for both samples, non-spherical particles with different sizes depending on BM time. $Gd_5Si_{1.3}Ge_{2.7}$ displays two transitions, due to an increase of the O(I) phase resulting in emergence of the second-order transition at higher temperatures. As a consequence, in $Gd_5Si_{1.3}Ge_{2.7}$, there is a decrease of the magnetocaloric effect with the increase of the milling time, which tends to stabilize after 90 min of ball milling. On the other hand, in the $Tb_5Si_2Ge_2$ case, the magnetization curves show that the BM process increases the temperature of structural transition, promoting its coupling with the magnetic one. As a result, the MCE increases with the milling time and at a linear rate, i.e., the MCE increases 23% after 150 min but does not show signals of stabilizing at least within the ball milling times studied here. Still the observed effects result from strain induced during the BM process. From this study we can conclude that BM, although a widely used and in principle simple technique, can be used as a tool to tune the MCE in $R_5(Si,Ge)_4$. The work also opens up an encouraging prospect for micro-scale magnetic refrigerators which would facilitate greater heat exchange times and thus higher frequency operation. This many also lead to higher efficiency, by potentially addressing one of the major limitation of the magnetic refrigeration at the present time. In the future, exploration of increased mill times for $Tb_5Si_2Ge_2$ composition will be performed to establish the optimum MCE achievable in this material system.

## ACKNOWLEDGMENTS

The authors acknowledge FCT for financial support through the projects: PTDC/CTM-NAN/115125/2009, CERN/FP/123585/2011, EXPL/EMS-ENE/2315/2013, FEDER/POCTIn0155/94 and the UK EPSRC EP/G060940/1. J.H. Belo thanks FCT for the Grant SFRH/BD/88440/2012. G.N.P. Oliveira gratefully thanks the FCT-Portugal for his FCT Grant No. SFRH/BD/80112/2012. A.M. Pereira and A.M.L. Lopes acknowledge the project NORTE-070124-FEDER-000070 for the financial support. A.L. Pires thanks for the Grant: PEst-OE/FIS/UI0275/2014 and Incentivo/FIS/UI0275/2014.

## REFERENCE

[1] Zimm C, Jastrab A, Sternberg A, Pecharsky V, Gschneidner K, Osborne M, et al. Advances in Cryogenic Engineering. Boston, MA: Springer US; 1998. doi:10.1007/978-1-4757-9047-4.

[2] Pecharsky VK, Gschneidner KA. Giant Magnetocaloric Effect in Gd5(Si2Ge2). Phys Rev Lett 1997;78:4494–7. doi:10.1103/PhysRevLett.78.4494.

[3]   Yu B., Gao Q, Zhang B, Meng X., Chen Z. Review on research of room temperature magnetic refrigeration. Int J Refrig 2003;26:622–36. doi:10.1016/S0140-7007(03)00048-3.

[4]   Gschneidner K a., Pecharsky VK. Thirty years of near room temperature magnetic cooling: Where we are today and future prospects. Int J Refrig 2008;31:945–61. doi:10.1016/j.ijrefrig.2008.01.004.

[5]   Zimm C, Boeder A, Chell J, Sternberg A, Fujita A, Fujieda S, et al. Design and performance of a permanent-magnet rotary refrigerator. Int J Refrig 2006;29:1302–6. doi:10.1016/j.ijrefrig.2006.07.014.

[6]   Romero Gómez J, Ferreiro Garcia R, De Miguel Catoira A, Romero Gómez M. Magnetocaloric effect: A review of the thermodynamic cycles in magnetic refrigeration. Renew Sustain Energy Rev 2013;17:74–82. doi:10.1016/j.rser.2012.09.027.

[7]   Gschneidner KA, Mudryk Y, Pecharsky VK. On the nature of the magnetocaloric effect of the first-order magnetostructural transition. Scr Mater 2012;67:572–7. doi:10.1016/j.scriptamat.2011.12.042.

[8]   Mudryk Y, Pecharsky VK, Gschneidner KA. Including Actinides. vol. 44. Elsevier; 2014. doi:10.1016/B978-0-444-62711-7.00262-0.

[9]   Fu H, Chen Y, Tu M, Zhang T. Phase analysis of Gd(SiGe) alloys prepared from different purity Gd with =0.475 and 0.43. Acta Mater 2005;53:2377–83. doi:10.1016/j.actamat.2005.01.045.

[10]  Fukamichi K, Fujita A, Fujieda S. Large magnetocaloric effects and thermal transport properties of La(FeSi)13 and their hydrides. J Alloys Compd 2006;408-412:307–12. doi:10.1016/j.jallcom.2005.04.022.

[11]  Hu FX, Ilyn M, Tishin AM, Sun JR, Wang GJ, Chen YF, et al. Direct measurements of magnetocaloric effect in the first-order system LaFe11.7Si1.3. J Appl Phys 2003;93:5503. doi:10.1063/1.1563036.

[12]  Liu J, Krautz M, Skokov K, Woodcock TG, Gutfleisch O. Systematic study of the microstructure, entropy change and adiabatic temperature change in optimized La–Fe–Si alloys. Acta Mater 2011;59:3602–11. doi:10.1016/j.actamat.2011.02.033.

[13]  Liu J, Gottschall T, Skokov KP, Moore JD, Gutfleisch O. Giant magnetocaloric effect driven by structural transitions. Nat Mater 2012;11:620–6. doi:10.1038/nmat3334.

[14]  Liu J, Woodcock TG, Scheerbaum N, Gutfleisch O. Influence of annealing on magnetic field-induced structural transformation and magnetocaloric effect in Ni–Mn–In–Co ribbons. Acta Mater 2009;57:4911–20. doi:10.1016/j.actamat.2009.06.054.

[15]  Pérez-Landazábal JI, Recarte V, Torrens-Serra J, Cesari E. Relaxation effects in magnetic-field-induced martensitic transformation of an Ni–Mn–In–Co alloy. Acta Mater 2014;71:117–25. doi:10.1016/j.actamat.2014.02.041.

[16]  Wada H, Tanabe Y. Giant magnetocaloric effect of MnAs(1−x)Sbx. Appl Phys Lett 2001;79:3302. doi:10.1063/1.1419048.

[17]  Aseguinolaza IR, Orue I, Svalov AV, Wilson K, Müllner P, Barandiarán JM, et al. Martensitic transformation in Ni–Mn–Ga/Si(100) thin films. Thin Solid Films 2014;558:449–54. doi:10.1016/j.tsf.2014.02.056.

[18]  Pires A, Belo J, Lopes A, Gomes I, Morellón L, Magen C, et al. Phase Competitions behind the Giant Magnetic Entropy Variation: Gd5Si2Ge2 and Tb5Si2Ge2 Case Studies. Entropy 2014;16:3813–31. doi:10.3390/e16073813.

[19]  Mohd Jani J, Leary M, Subic A, Gibson MA. A review of shape memory alloy research, applications and opportunities. Mater Des 2014;56:1078–113. doi:10.1016/j.matdes.2013.11.084.

[20]  Zhang TB, Provenzano V, Chen YG, Shull RD. Magnetic properties of a high energy ball-milled amorphous Gd5Si1.8Ge1.8Sn0.4 alloy. Solid State Commun 2008;147:107–10. doi:10.1016/j.ssc.2008.05.009.

[21]  Miller CW, Williams D V., Bingham NS, Srikanth H. Magnetocaloric effect in Gd/W thin film heterostructures. J Appl Phys 2010;107:09A903. doi:10.1063/1.3335515.

[22]  Miller CW, Belyea DD, Kirby BJ. Magnetocaloric effect in nanoscale thin films and heterostructures. J Vac Sci Technol A Vacuum, Surfaces, Film 2014;32:040802. doi:10.1116/1.4882858.

[23]  Lyubina J, Gutfleisch O, Kuz'min MD, Richter M. La(Fe,Si)13-based magnetic refrigerants obtained by novel processing routes. J Magn Magn Mater 2009;321:3571–7. doi:10.1016/j.jmmm.2008.03.063.

[24]  Lovell E, Pereira AM, Caplin AD, Lyubina J, Cohen LF. Dynamics of the First-Order Metamagnetic Transition in Magnetocaloric La(Fe,Si) 13 : Reducing Hysteresis. Adv Energy Mater 2014:n/a – n/a. doi:10.1002/aenm.201401639.

[25] Gomes MB, de Oliveira NA. Magnetocaloric effect in Tb5Si2Ge2 under applied pressure. J Magn Magn Mater 2008;320:e153–5. doi:10.1016/j.jmmm.2008.02.038.

[26] Trevizoli PV, Alves CS, Mendes M a. B, Carvalho a. MG, Gama S. Powder metallurgy influences on the magnetic properties of Gd5.09Ge2.03Si1.88 alloy. J Magn Magn Mater 2008;320:1582–5. doi:10.1016/j.jmmm.2008.01.007.

[27] Moya X, Kar-Narayan S, Mathur ND. Caloric materials near ferroic phase transitions. Nat Mater 2014;13:439–50. doi:10.1038/nmat3951.

[28] Khovaylo V V, Rodionova V V, Shevyrtalov SN, Novosad V. Magnetocaloric effect in "reduced" dimensions: Thin films, ribbons, and microwires of Heusler alloys and related compounds. Phys Status Solidi 2014;251:2104–13. doi:10.1002/pssb.201451217.

[29] Kumaresavanji M, Sousa CT, Pires A, Pereira AM, Lopes AML, Araujo JP. Room temperature magnetocaloric effect and refrigerant capacitance in La0.7Sr0.3MnO3 nanotube arrays. Appl Phys Lett 2014;105:083110. doi:10.1063/1.4894175.

[30] Svalov A V., Vas'kovskiy VO, Schegoleva NN, Kurlyandskaya G V. Effect of the layer thickness on the magnetic properties and structure of terbium in (Tb/Ti)n and (Tb/Si)n multilayer films. Tech Phys 2013;50:914–7. doi:10.1134/1.1994973.

[31] Suryanarayana C. Mechanical alloying and milling. Prog Mater Sci 2001;46:1–184. doi:10.1016/S0079-6425(99)00010-9.

[32] Sharifati A, Sharafi S. Structural and magnetic properties of nanostructured (Fe70Co30)100−xCux alloy prepared by high energy ball milling. Mater Des 2012;41:8–15. doi:10.1016/j.matdes.2012.04.047.

[33] Rajkumar DM, Manivel Raja M, Gopalan R, Chandrasekaran V. Magnetocaloric effect in high-energy ball-milled Gd5Si2Ge2 and Gd5Si2Ge2/Fe nanopowders. J Magn Magn Mater 2008;320:1479–84. doi:10.1016/j.jmmm.2007.12.005.

[34] Phan MH, Morales MB, Chinnasamy CN, Latha B, Harris VG, Srikanth H. Magnetocaloric effect in bulk and nanostructured Gd3Fe5O12 materials. J Phys D Appl Phys 2009;42:115007. doi:10.1088/0022-3727/42/11/115007.

[35] Kaya M, Rezaeivala M, Yüzüak E, Akturk S, Dincer I, Elerman Y. Effects of size reduction on the magnetic and magnetocaloric properties of NdMn 2 Ge 2 nanoparticles prepared by high-energy ball milling. Phys Status Solidi 2015;252:192–7. doi:10.1002/pssb.201451127.

[36] Biswas A, Chandra S, Stefanoski S, Blázquez JS, Ipus JJ, Conde A, et al. Enhanced cryogenic magnetocaloric effect in Eu8Ga16Ge30 clathrate nanocrystals. J Appl Phys 2015;117:033903. doi:10.1063/1.4906280.

[37] Giovanna do Couto G, Svitlyk V, Jafelicci M, Mozharivskyj Y. Bulk and high-energy ball-milled Gd5Si2Ge2: Comparative study of magnetic and magnetocaloric properties. Solid State Sci 2011;13:209–15. doi:10.1016/j.solidstatesciences.2010.11.016.

[38] Ritter C. Magnetic and structural phase diagram of Tb5(SixGe1-x)4. Phys Rev B 2002;65:094405. doi:10.1103/PhysRevB.65.094405.

[39] Pereira AM, Magen C, Braga ME, Pinto RP, Fermento R, Algarabel PA, et al. Transport properties near the magneto/structural transition of Tb5Si2Ge2. J Non Cryst Solids 2008;354:5298–300. doi:10.1016/j.jnoncrysol.2008.05.069.

[40] Juan Y, Kaxiras E. High-pressure plastic flow in silicon: A first-principles theoretical study. J Comput Mater Des 1993;1:55–62. doi:10.1007/BF00712816.

[41] Caron L, Ou ZQ, Nguyen TT, Cam Thanh DT, Tegus O, Brück E. On the determination of the magnetic entropy change in materials with first-order transitions. J Magn Magn Mater 2009;321:3559–66. doi:10.1016/j.jmmm.2009.06.086.

[42] Pecharsky VK, Gschneidner KA. Phase relationships and crystallography in the pseudobinary system Gd5Si4-Gd5Ge4. J Alloys Compd 1997;260:98–106. doi:10.1016/S0925-8388(97)00143-6.

[43] Babizhetskyy V, Roger J, Députier S, Jardin R, Bauer J, Guérin R. Solid state phase equilibria in the Gd–Si–B system at 1270K. J Solid State Chem 2004;177:415–24. doi:10.1016/j.jssc.2003.02.011.

[44] Llamazares JLS, Pérez MJ, Álvarez P, Santos JD, Sánchez ML, Hernando B, et al. The effect of ball milling in the microstructure and magnetic properties of Pr2Fe17 compound. J Alloys Compd 2009;483:682–5. doi:10.1016/j.jallcom.2008.07.210.

[45] Gschneidner KA, Pecharsky VK, Tsokol AO. Recent developments in magnetocaloric materials. Reports Prog Phys 2005;68:1479–539. doi:10.1088/0034-4885/68/6/R04.

[46] Morellon L, Ritter C, Magen C, Algarabel P, Ibarra M. Magnetic-martensitic transition of Tb5Si2Ge2 studied with neutron powder diffraction. Phys Rev B 2003;68:024417. doi:10.1103/PhysRevB.68.024417.

[47] Morellon L, Arnold Z, Magen C, Ritter C, Prokhnenko O, Skorokhod Y, et al. Pressure Enhancement of the Giant Magnetocaloric Effect in Tb5Si2Ge2. Phys Rev Lett 2004;93:137201. doi:10.1103/PhysRevLett.93.137201.

[48] Provenzano V, Shapiro AJ, Shull RD. Reduction of hysteresis losses in the magnetic refrigerant Gd5Ge2Si2 by the addition of iron. Nature 2004;429:853–7. doi:10.1038/nature02657.

[49] Belo JH, Pereira AM, Araújo JP, de la Cruz C, dos Santos AM, Gonçalves JN, et al. Tailoring the magnetism of Tb5Si2Ge2 compounds by La substitution. Phys Rev B 2012;86:014403. doi:10.1103/PhysRevB.86.014403.

[50] Pereira AM, dos Santos AM, Magen C, Sousa JB, Algarabel PA, Ren Y, et al. Understanding the role played by Fe on the tuning of magnetocaloric effect in Tb5Si2Ge2. Appl Phys Lett 2011;98:122501. doi:10.1063/1.3567920.

[51] Moreno-Ramírez LM, Ipus JJ, Franco V, Blázquez JS, Conde A. Analysis of magnetocaloric effect of ball milled amorphous alloys: Demagnetizing factor and Curie temperature distribution. J Alloys Compd 2015;622:606–9. doi:10.1016/j.jallcom.2014.10.134.